\documentclass[12pt,a4paper,numbers, numbers, sort&compress, draftclsnofoot, onecolumn, 12pt]{IEEEtran}
\usepackage[latin9]{inputenc}
\usepackage{color}
\usepackage{array}
\usepackage{multirow}
\usepackage{amsmath}
\usepackage{amssymb}
\usepackage{graphicx}
\usepackage[numbers]{natbib}

\makeatletter

\pdfpageheight\paperheight
\pdfpagewidth\paperwidth

\providecommand{\tabularnewline}{\\}
\def\@tempa{12}\ifx\@ptsize\@tempa
\def\@normalsize{\@setsize\normalsize{14pt}\xiipt\@xiipt
\abovedisplayskip 2pt plus2pt minus2pt\belowdisplayskip \abovedisplayskip
\abovedisplayshortskip \z@ plus3pt\belowdisplayshortskip .2pt plus2pt minus2pt}
\def\small{\@setsize\small{11.4pt}\xpt\@xpt}
\def\footnotesize{\@setsize\footnotesize{10pt}\ixpt\@ixpt}
\def\scriptsize{\@setsize\scriptsize{9pt}\viiipt\@viiipt}
\def\tiny{\@setsize\tiny{8pt}\vipt\@vipt}
\def\large{\@setsize\large{18pt}\xivpt\@xivpt}
\def\Large{\@setsize\Large{22pt}\xviipt\@xviipt}
\def\LARGE{\@setsize\LARGE{25pt}\xxpt\@xxpt}
\def\huge{\@setsize\huge{30pt}\xxvpt\@xxvpt}

\fi

\def\@tempa{12}\ifx\@ptsize\@tempa
\def\@normalsize{\@setsize\normalsize{14pt}\xiipt\@xiipt
\abovedisplayskip 2pt plus2pt minus2pt\belowdisplayskip \abovedisplayskip
\abovedisplayshortskip \z@ plus3pt\belowdisplayshortskip .2pt plus2pt minus2pt}
\def\small{\@setsize\small{11.4pt}\xpt\@xpt}
\def\footnotesize{\@setsize\footnotesize{10pt}\ixpt\@ixpt}
\def\scriptsize{\@setsize\scriptsize{9pt}\viiipt\@viiipt}
\def\tiny{\@setsize\tiny{8pt}\vipt\@vipt}
\def\large{\@setsize\large{18pt}\xivpt\@xivpt}
\def\Large{\@setsize\Large{22pt}\xviipt\@xviipt}
\def\LARGE{\@setsize\LARGE{25pt}\xxpt\@xxpt}
\def\huge{\@setsize\huge{30pt}\xxvpt\@xxvpt}

\fi
\usepackage{url}

\usepackage{graphicx}

\usepackage{lineno}\usepackage{bm}
\usepackage{subfig}%\usepackage[linesnumbered]{algorithm2e}
\usepackage{algorithm}\usepackage{algorithmic} %Use Input in the format of Algorithm
\providecommand{\tabularnewline}{\\}

\@ifundefined{showcaptionsetup}{}{%
 \PassOptionsToPackage{caption=false}{subfig}}
\usepackage{subfig}
\makeatother

\begin{document}

\title{{\huge {Energy Harvesting Small Cell Networks: Feasibility, Deployment
and Operation}}}

\author{\IEEEauthorblockN{Yuyi Mao, Yaming Luo, Jun Zhang, Khaled B. Letaief,
\emph{Fellow, IEEE}\\
 } \IEEEauthorblockA{Hong Kong University of Science and Technology}
\textit{}%
\thanks{This work is supported by the Hong Kong Research Grant Council under
Grant No. 610212.%
}}

\maketitle
%\boldmath

\begin{abstract}
Small cell networks (SCNs) have attracted great attention in recent
years due to their potential to meet the exponential growth of mobile
data traffic and the increasing demand for better quality of service
and user experience in mobile applications. Nevertheless, a wide deployment
of SCNs has not happened yet because of the complexity in the network
planning and optimization, as well as the high expenditure involved
in deployment and operation. In particular, it is difficult to provide
grid power supply to all the small cell base stations (SCBSs) in a
cost effective way. Moreover, a dense deployment of SCBSs, which is
needed to meet the capacity and coverage of the next generation wireless
networks, will increase operators' electricity bills and lead to significant
carbon emission. Thus, it is crucial to exploit off-grid and green
energy sources to power SCNs, for which energy harvesting (EH) technology
is a viable solution. In this article, we will conduct a comprehensive
study of EH-SCNs, and investigate important aspects, including the
feasibility analysis, network deployment, and network operation issues.
The advantages, as well as unique challenges, of EH-SCNs will be highlighted,
together with potential solutions and effective design methodologies.

\thispagestyle{empty}
\setcounter{page}{0}

\newpage{}
\end{abstract}

%\begin{keywords}
%Energy Harvesting, Small  Cell Association, Power Allocation
%\end{keywords}
% \fi
\IEEEpeerreviewmaketitle

\section{Introduction}

The proliferation of mobile devices, such as smart phones and tablets,
is boosting the data traffic explosion in wireless ecosystems. In
this context, cellular networks are faced with the challenges of providing
enormous network capacity, achieving superior cellular coverage, and
improving users' quality of experience. The small cell network (SCN)
is a cost-effective and energy-efficient network paradigm to tackle
these challenges. In SCNs, the densely-deployed small cell base stations
(SCBSs), including micro, pico, and femto-cells, bring the spatial
reuse of radio resources to a new level, which will then help improve
the area spectral efficiency and user experience. Besides this, the
low-cost and low-power SCBSs can be easily installed without costly
cell site acquisition, and their self-organization manner further
helps save operating expenditures \citep{Chandrasekhar08,Hoydis1103}.

However, as the SCBSs are densely and irregularly located, some of
them may be inaccessible to the power grid. Moreover, the network
power consumption of the SCNs will be high despite the small power
consumption of a single SCBS, which will produce a significant amount
of carbon emissions. As a result, it is desirable to exploit off-grid
and green energy sources to power the SCNs. Energy harvesting (EH)
technology is a viable and promising solution, which can harvest ambient
renewable energy, e.g., solar and wind energy, to power SCBSs \citep{THan1312}.
It is estimated that applying EH techniques to SCNs can achieve a
20\% $CO_{2}$ reduction in the information and communication technology
(ICT) industry \citep{Piro1301}.

Communication networks with EH capability have been extensively studied
in recent years, from point-to-point systems \citep{Gunduz1401,CKHO1209},
to multi-user systems \citep{JYang1202} and EH heterogeneous networks
\citep{Dhillon1405}. However, so far, there has been no systematic
study on how to effectively utilize the EH techniques in SCNs, i.e.,
how to power the SCBSs by EH, how to deploy EH-SCBSs, and how to optimize
the network operations for EH-SCNs. The goal of this article is to
provide a comprehensive study for EH-SCNs. Specifically, the feasibility
analysis of EH-SCNs will be conducted first, and then the network
deployment issues will be addressed from the basic tradeoffs to practical
deployment considerations. Over the deployed EH-SCNs, the challenges
and design methodologies for network operation will be elaborated.

\section{Powering Small Cell Networks by Energy Harvesting - A Feasibility
Analysis}

In this section, we will investigate the feasibility of powering SCNs
by renewable energy sources. We will first highlight the main differences
between the energy consumption models for the macro base station (BS)
and SCBSs, following which the potential of different EH techniques
to power SCBSs will be discussed. In particular, it will be revealed
that the hybrid solar-wind energy harvester will be an ideal candidate
to enable EH-SCNs.

\subsection{The Feasibility of EH-SCNs}

The energy consumption models of SCBSs are fundamentally different
from the macro BS, which are specified as follows:
\begin{itemize}
\item The communication distances from the SCBSs to mobile users will be
significantly reduced compared to macro BSs, as SCBSs will be densely
deployed \citep{Hoydis1103}. Consequently, the transmit power of
SCBSs will be greatly reduced. For example, the maximum transmit power
of a typical femto BS is 17 dBm, compared to 43 dBm for a macro BS
\citep{Auer1110}.
\item The baseband processing in SCBSs is much simpler than in the macro
BS, since several key operations are eliminated, such as the digital
pre-distortion.
\item Cooling in the macro BS accounts for around 10\% of the total BS power
consumption \citep{Auer1110}, while SCBSs can be cooled by natural
air circulation.
\end{itemize}
Given the above discussion, the power consumption of an SCBS is orders
of magnitude smaller than the typical macro BS. Specifically, the
power consumption of a typical macro BS is $225$ W per transceiver.
In contrast, it is $72.3$ W/$7.3$ W/$5.2$ W for a micro/pico/femto
BS \citep{Auer1110}. Thus, it will be more feasible to power SCBSs
by EH. On the other hand, as SCBSs need to be densely deployed, their
total energy consumption may still be high. Thus, powering SCBSs by
renewable energy sources is also motivated by environmental concern.

\begin{table}
\center \protect

\caption{Existing Energy Harvesting Techniques}

\begin{tabular}{|c|c|c|c|c|}
\hline
\multirow{2}{*}{\textbf{Energy Sources}} & \multirow{2}{*}{\textbf{Characteristics}} & \textbf{Implementation } & \textbf{Amount of } & \multirow{2}{*}{\textbf{Typical Applications}}\tabularnewline
 &  & \textbf{Techniques} & \textbf{Harvested Energy } & \tabularnewline
\hline
\hline
\multirow{2}{*}{Solar \citep{Sudevalayam1109}} & Uncontrollable,  & \multirow{2}{*}{Photovoltaic cells } & \multirow{2}{*}{15 $\mathrm{mW/cm^{2}}$ } & Wireless sensor, \tabularnewline
 & predictable &  &  & household appliances\tabularnewline
\hline
\multirow{2}{*}{Wind \citep{Sudevalayam1109} } & Uncontrollable,  & \multirow{2}{*}{Anemometer} & 85 W (rotor diameter  & Wireless sensor, \tabularnewline
 & predictable &  & 1m, wind speed 8m/s) & household appliances\tabularnewline
\hline
Environmental & Uncontrollable,  & \multirow{2}{*}{Electromagnetic induction } & \multirow{2}{*}{0.2 $\mathrm{mW/cm^{2}}$ } & Wireless sensor, \tabularnewline
Vibration \citep{Sudevalayam1109} & unpredictable &  &  & consumer electronic\tabularnewline
\hline
\multirow{2}{*}{Human Motion \citep{Sudevalayam1109} } & Controllable,  & \multirow{2}{*}{Piezoelectric} & Finger motion: 2.1 mW & On-body monitoring, \tabularnewline
 & predictable &  & footfalls: 5 W & portable devices\tabularnewline
\hline
\multirow{2}{*}{Thermal \citep{Lu1010}} & Uncontrollable,  & \multirow{2}{*}{Thermopiles} & \multirow{2}{*}{$\approx40$ mW } & \multirow{2}{*}{Wireless sensor }\tabularnewline
 & unpredictable &  &  & \tabularnewline
\hline
Ambient & Uncontrollable,  & \multirow{2}{*}{Rectification \& filtering } & \multirow{2}{*}{$<$0.2 mW } & \multirow{2}{*}{RFID, low power device }\tabularnewline
RF Signal \citep{Lu1406} & unpredictable &  &  & \tabularnewline
\hline
\multirow{2}{*}{Biomass \citep{Huang1309} } & Controllable,  & \multirow{2}{*}{Microbial fuel cells } & \multirow{2}{*}{153 $\mathrm{mW/m^{2}}$} & \multirow{2}{*}{Underwater sensor}\tabularnewline
 & predictable &  &  & \tabularnewline
\hline
\end{tabular}
\end{table}

To check the feasibility of powering SCNs via energy harvesting, we
summarize the main energy sources of EH techniques in Table I. We
see that most of the existing applications of EH techniques are limited
to low-power electronic devices, mainly due to the low EH rates. Considering
the typical power consumption of an SCBS ($>$5W), only a few of the
EH sources are applicable, among which solar and wind energy are the
two most promising ones due to the following reasons:
\begin{itemize}
\item Sufficient harvested energy can be guaranteed with either a solar
or wind energy harvester. For example, $100$ W electric power can
be generated by either a 121 cm $\times$ 53.6 cm solar panel under
rated sunlight radiation, or by a rotor with a 1 m diameter under
an 8 m/s wind speed.
\item Such energy harvesters are cost-effective, due to their decades of
lifetime and almost negligible maintenance expenditure. The main cost
of solar/wind energy harvesters originates from the deployment stage,
which has been decreasing dramatically in recent years.
\item Many industrial companies are actively participating in developing
solar and wind energy harvesters, e.g., Suntech, First Solar, Sunpower,
and Trina Solar for solar energy, or GE Energy, Vestas, Siemens Wind
Power, and Goldwind for wind energy.
\end{itemize}

Though solar and wind harvesters enjoy high harvesting rates and low
cost, the time variation of the energy source poses challenges to
solar/wind power generation. Fortunately it turns out that solar and
wind are a good complement to each other. On daily timescales, high
pressure areas tend to bring clear skies and low surface winds, which
is favorable for solar harvesters, whereas low pressure areas tend
to be windier and cloudier, and thus are good for wind harvesters.
On seasonal timescales, solar energy peaks in summer, whereas in many
areas wind energy is lower in summer and higher in winter. We will
demonstrate such a complementary effect in the following case study.

\textbf{\emph{A case study:}} We will use real solar and wind power
generation data by the Elia Group in Belgium\footnote{Elia, Power generation, available online at http://www.elia.be/en/grid-data/power-generation.}.
The normalized energy profiles on daily timescales are shown first.
Based on the measured data from 0:00 am, 15 June to 0:00 am, 17 June,
2014, the average solar/wind power is shown in Fig. 1 (a), where the
EH rates are sampled (averaged) every 15 minutes. We see that the
peak of the solar power always coincides with the valley of the wind
power, and vice versa. Next we show the energy profiles on seasonal
timescales. Based on the data from 0:00 am, 17 May 2013 to 0:00 am,
02 July 2014, the average solar/wind power, averaged every 15 days,
is shown in Fig. 1 (b). We see that the solar power achieves its peak
during June-August, while the wind power reaches its bottom. An opposite
trend is observed during December-February.

This case study reveals that a combination of solar and wind energy
is a good candidate for the energy source of SCBSs. Actually, BSs
powered by hybrid solar-wind energy have already drawn great attention
from the industry. For example, the Turkish mobile operator Avea,
and the leading equipment vendor Huawei have shown great interest
in such BSs. Particularly, Wind-Fi, a renewable energy BS designed
by the Centre for White Space Communications, enables wireless networks
to operate entirely on solar and wind energy, and achieves 99.98\%
reliability\footnote{Centre for White Space Communications, WindFi: Renewable-Energy Wireless Basestations, available online at http://www.wirelesswhitespace.org/projects/wind-fi-renewable-energy-basestation.aspx.}.

\subsection{The Prospect of EH-SCNs}

The above discussion demonstrates that EH technology, particularly
solar and wind harvesters, is a viable green energy solution for SCNs.
Therefore, the EH-SCNs in the near future may consist of solar-powered
SCBSs, wind-powered SCBSs, hybrid solar-wind-powered SCBSs, EH-SCBSs
powered by other energy sources, and conventional grid-powered SCBSs
as well. EH-SCNs will not only reduce the deployment cost and energy
bills for the operators, but will also be more environmentally friendly
and thus can enable sustainable growth of wireless networks. An example
of such a network is shown in Fig. 2.

Powering SCNs by EH sources will bring new design challenges. The
network coverage and the operating reliability will be difficult to
guarantee since harvested energy varies daily and seasonally. Adjustments
in the communication protocols and transmission strategies will be
needed. In the following part of this paper, special attention will
be paid towards addressing the upcoming design issues in network deployment
as well as in network operation.

\section{Network Deployment of EH-SCNs}

Network deployment is the first step towards designing an effective
EH-SCN. A key question to ask is how many EH-SCBSs are needed? Thus,
in this section we will investigate the impact of the SCBS density
on network performance and cost, which will reveal some interesting
tradeoffs in EH-SCNs. Other deployment issues will then be discussed.

\subsection{Basic Tradeoffs }

The density of an EH-SCN will determine its performance, as well as
the network cost. Increasing the EH-SCBS density can improve the coverage
and throughput, but it will also increase the deployment cost. On
the other hand, with a low density of EH-SCBSs, more grid-powered
SCBSs will be needed to maintain the coverage, which will consume
more nonrenewable energy and increase the energy bills. In the following,
we will provide simulation results to illustrate these tradeoffs.
The outage probability will be adopted as the performance metric,
which is the portion of users that cannot be successfully served.

\textbf{\emph{1) Tradeoff Between Outage Probability and EH-SCBS Density:}}
We will first consider to provide network coverage only with off-grid
EH-SCBSs, i.e., without any support of the grid. We assume each user
is associated with its nearest SCBS, and we will ignore co-channel
interference as the main purpose is to guarantee network coverage.
For each SCBS, energy arrives intermittently with an average EH rate
$P_{EH}$. In each time slot, part of the harvested energy will be
used to serve its users, while the remaining part will be stored in
a battery with capacity $C_{B}$. To investigate the impact of $C_{B}$,
we consider two extreme cases, i.e., $C_{B}=0$ or $C_{B}=\infty$.
The transmit power for each user is determined to satisfy its receive
SNR requirement $\gamma_{th}$. We ignore the circuit power consumption
of the SCBSs unless otherwise mentioned. Each SCBS will serve all
of its associated users if the available energy is sufficient; otherwise,
it maximizes the number of served users. The SCBS and user densities
are denoted by $\lambda_{BS}$ and $\lambda_{u}$, respectively. The
tradeoff between the outage probability $p_{out}$ and $\lambda_{BS}$
is shown in Fig. 3 (a). Key observations can be drawn:
\begin{itemize}
\item The outage probability decreases with $\lambda_{BS}$, but the decreasing
rate reduces as $\lambda_{BS}$ increases further. Specifically, to
achieve $p_{out}=10\%$ in this EH-SCN with $\lambda_{u}=10^{-3}\textrm{ m}^{-2}$,
we can deploy SCBSs with density $\lambda_{BS}\geq1.7\times10^{-4}\textrm{ m}^{-2}$
if each SCBS is with $P_{EH}=20\textrm{ mW}$ when supported by a
battery with large enough capacity, or with $\lambda_{BS}\geq2.1\times10^{-4}\textrm{ m}^{-2}$
if each is without a battery. Both are within the typical network
density range of the SCNs \citep{Chang1405}.
\item The battery capacity has little influence on $p_{out}$ when $\lambda_{BS}$
is either very small or very large. When $\lambda_{BS}$ is very small,
as the harvested energy is insufficient almost all the time, the energy
will be exhausted immediately after it arrives. On the other hand,
if $\lambda_{BS}$ is very large, the current harvested energy will
be more than enough, and there is no need to consume the energy in
the battery.
\item Increasing either $P_{EH}$ or $\lambda_{BS}$ will reduce $p_{out}$.
Meanwhile, interestingly, increasing $\lambda_{BS}$ brings more performance
improvement, which can be explained intuitively. Doubling $\lambda_{BS}$
not only doubles the available energy in the whole network, but also
reduces the transmission distances on average.
\end{itemize}

\textbf{\emph{2) Tradeoff Between Grid Power Consumption and EH-SCBS
Density:}} In this part, we will assume that all the SCBSs are on-grid
SCBSs, that is, the power grid is retained as the backup energy source
for each SCBS. With a stable power supply, it is easy to guarantee
coverage, and thus the focus is on the impact of the EH-SCBS density
on the grid power consumption. At each SCBS, the harvested energy
will be exhausted first, and the grid power will be used only when
necessary. The tradeoff between the grid power consumption $P_{G}$
and $\lambda_{BS}$ is shown in Fig. 3 (b). Key observations can be
drawn:
\begin{itemize}
\item The grid power consumption $P_{G}$ decreases with the SCBS density
$\lambda_{BS}$, and the optimal EH-SCBS density can be chosen to
minimize the network deployment and operating expenditures. For example,
assume the electricity price is \$0.1971/$\textrm{KW}\textrm{h}$,
while each SCBS costs \$135, of which \$35 is for the 10 W photovoltaic
cell\footnote{This value takes the circuit power consumption of an EH-SCBS into account.},
and \$100 is for the BS equipment. Normally, the lifetime of EH-SCNs
is around 10 years. Then, if $\lambda_{u}=3\times10^{-3}\textrm{ m}^{-2}$
and $P_{EH}=20\textrm{ mW}$, we can find the optimal SCBS density
as $7.5\times10^{-5}\textrm{ m}^{-2}$.
\item To reduce $P_{G}$, increasing $\lambda_{BS}$ is more effective than
increasing $P_{EH}$. The impact of $C_{B}$ on $P_{G}$ is negligible
when $\lambda_{BS}$ is extremely small or extremely large.
\end{itemize}

\textbf{\emph{3) Tradeoff Between Outage Probability and Grid Power
Consumption:}} Compared to off-grid SCBSs, on-grid SCBSs make it easy
to guarantee coverage with a stable power supply, but they are more
difficult to deploy due to the grid power supply, and they will also
increase the non-renewable energy consumption. In this part, we will
consider an SCN with both off-grid and on-grid SCBSs, while the total
density is fixed. By varying the density of off-grid SCBSs, we can
achieve different tradeoffs between the outage probability and the
grid power consumption. With the ratio of on-grid SCBSs, denoted as
$\eta$, increasing from 0 to 1, the outage probability will decrease,
while the grid power consumption will increase. The relationship between
$p_{out}$ and $P_{G}$ is shown in Fig. 3 (c), from which we can
make the following observations:
\begin{itemize}
\item Changing $\eta$ can adjust the tradeoff between the outage probability
and the grid power consumption. For example, when $\lambda_{BS}=2\times10^{-4}\textrm{ m}^{-2}$,
$P_{EH}=40\textrm{ mW}$, we can achieve the outage probability $p_{out}=0$
with $P_{G}=2.38\textrm{ W}$ by setting $\eta=1$. Alternatively,
we can achieve $p_{out}=0.08$ with $P_{G}=1.46$ W by setting $\eta=0.6$,
i.e., replacing 40\% of the on-grid SCBSs with off-grid SCBSs, we
can reduce the grid power consumption by $\sim$40\% with a slight
performance degradation.
\item The outage probability scales linearly with $P_{G}$, as both $p_{out}$
and $P_{G}$ scale linearly with $\eta$ due to the independent and
identical settings for different SCBSs, such as their locations and
EH rates.
\end{itemize}

\subsection{Deployment Issues}

The previous discussions on the three basic tradeoffs in EH-SCNs provide
us with the following deployment guidelines:
\begin{itemize}
\item Satisfactory coverage can be guaranteed in EH-SCNs with a reasonable
network density. By carefully determining the network density, we
can not only balance between network performance and deployment cost,
but also achieve a tradeoff between performance and grid power consumption.
\item To improve the network performance or to save the grid power consumption,
it is more effective to increase the SCBS density than to increase
the EH rate of each SCBS (e.g., by deploying a larger solar panel).
\item When the EH-SCBS density is extremely small or extremely large, battery
capacity has little influence on the network performance or the grid
power consumption.
\end{itemize}

So far, the considered scenarios are rather simplified. For example,
co-channel interference between users is ignored. The BS power consumption
model is also ideal, as only the transmit power is considered, while
in practice, for a femto BS, when the transmit power is 25 mW, around
5.2 W is consumed by the whole BS \citep{Auer1110}. Therefore, a
more detailed investigation will be needed. One useful tool for network
deployment is the spatial network model, as adopted in \citep{Dhillon1405}
and \citep{Chang1405}. Such a network model can help to provide analytical
results for performance evaluation, which may then provide guidelines
for network deployment and avoid the time-consuming simulations. Moreover,
we need to take realistic physical and social factors into consideration.
Generally, BS locations can be adapted to the spatial traffic profile,
i.e., more SCBSs should be deployed in the traffic hotspots to meet
the high communication demand. Moreover, for a given location, the
EH source should be chosen according to the ambient energy availability
and their economic costs. For instance, a wind-powered SCBS is preferred
to solar-powered SCBS at the seashore due to the abundant amount and
the installation convenience of wind energy.

\section{Network Operation of EH-SCNs}

\textcolor{black}{In the last section, we investigated the deployment
issues in EH-SCNs with simplified network operations. In a practically
deployed EH-SCN, the network operation should be carefully designed
to optimize the network performance. Due to the spatial and temporal
variations of the EH conditions, the network operation strategies
for conventional grid-powered SCNs are no longer applicable to EH-SCNs.
In this section, with the joint power assignment and cell association
problem as an example, we will illustrate the unique design challenges
and some promising methodologies for EH-SCN network operations.}

\subsection{\textcolor{black}{Power Assignment and }Cell Association in EH-SCNs}

Introducing EH-SCBSs will bring unique challenges for the SCBS power
assignment and cell association problem, i.e., determining which SCBS
each mobile should be associated with, and at which power level each
SCBS should choose to transmit the signal. In particular, the following
aspects should be considered:
\begin{itemize}
\item \emph{To incorporate the temporal and spatial variation of the available
energy.} In conventional SCNs, fixed cell association is normally
adopted, e.g., the users are associated with their nearest SCBSs \citep{Chang1405}.
In SCNs, the temporal and spatial variation of the available EH source
makes fixed association inapplicable, and a given user will need to
be associated with different SCBSs during different periods. Thus,
the design of cell association policies should balance the energy
utilization of different SCBSs.
\item \emph{To incorporate the coupling among different users/SCBSs.} For
a given SCBS, if it allocates a too high transmit power to serve one
user, it may easily exhaust its available energy and may not be able
to serve other users. Thus some of its users need to be offloaded
to other SCBSs, the available energy of which may be quickly depleted.
This coupling among users/SCBSs renders power assignment of each SCBS
and cell association quite complicated in EH-SCNs.
\end{itemize}
To demonstrate these aspects in more detail, we will next consider
two specific design problems.

\textbf{\emph{1) Performance Optimization for Off-grid EH-SCNs:}}
We first consider an EH-SCN with $M$ off-grid EH SCBSs and $K$ mobile
users, where each user is served by one SCBS in each time slot. To
provide satisfactory performance to these users, an efficient joint
cell association and power assignment policy should be developed.
For simplicity, we assume a constant EH rate for each SCBS, but different
SCBSs may have different EH rates. The design objective is to maximize
the minimum average signal-to-noise ratio (SNR) among the $K$ users,
and thus, fair performance can be provided. This problem can be shown
to be NP-hard. To reduce the computational complexity, we propose
a low-complexity sub-optimal solution based on the threshold-bisection
algorithm proposed in \citep{Luo1404}. The proposed method will not
only balance the energy usage at different SCBSs, but also take the
future available energy at each SCBS into consideration, i.e., it
considers both spatial and temporal energy variation.

To show the effectiveness of the proposed method, we introduce a performance
upper bound and two baseline policies. The upper bound is obtained
by allowing multiple SCBSs to jointly serve all the users using distributed
beamforming, denoted as `distributed BF'. The first baseline policy
adopts distance-based cell association, where each user is served
by its nearest SCBS. The second one adopts SNR-based cell association,
where each user will be associated to the SCBS that provides the highest
receive SNR with the available energy. The performances of different
policies are shown in Fig. 4. Key observations can be drawn:
\begin{itemize}
\item The proposed solution greatly outperforms both baseline policies and
achieves performance close to the upper bound.
\item The distance-based policy suffers performance loss as it neglects
the spatial variation of available energy at different SCBSs. Therefore,
conventional cell association strategies cannot be directly adopted
in EH-SCNs.
\item The SNR-based policy performs better than the distance-based policy,
as it utilizes information on both the distance and the current energy
state. However, it still suffers performance degradation as it makes
decisions based only on the current system state and neglects the
coupling in different transmission blocks as well as among different
users.
\end{itemize}

In summary, the cell association policies should be redesigned for
EH-SCNs and important aspects should be taken into consideration,
including the temporal and spatial variation of the energy, and the
coupling among SCBSs/mobile users. It is difficult to obtain optimal
solutions, but effective sub-optimal solutions can be developed by
considering the unique properties of EH-SCNs.

\textbf{\emph{2) Grid Power Minimization for On-grid EH-SCBSs:}} In
this part, we will consider an SCN consisting of both EH-SCBSs and
grid-powered SCBSs. The design objective is to minimize the power
consumption of grid-powered SCBSs by adaptive cell association. For
simplicity, we will assume only one of the $M$ SCBSs is powered by
the grid, and we focus on the single-user case. All other assumptions
are the same as the previous design problem. For this problem, we
have obtained the following two optimal transmission strategies:
\begin{itemize}
\item \textbf{The Save-Transmit Strategy:} For this solution, there exists
a critical time slot, before which the user is served by the grid-powered
SCBS, while afterwards, the EH-SCBSs take turns to serve the user.
This strategy reflects an innate characteristic of EH systems, i.e.,
with a given number of time slots to use EH-SCBSs, deferring these
time slots will not deteriorate the performance. However, non-causal
EH information is required to obtain the critical time slot index.
\item \textbf{The Greedy-Transmit Strategy: }For this solution, in each
time slot, if possible, the user will be served by one of the EH-SCBSs
that has enough energy. Otherwise, the user is served by the grid-powered
SCBS. This solution is extremely simple, as the decision in each time
slot only depends on the current energy state of each EH-SCBS, irrespective
of the future EH information.
\end{itemize}

We illustrate these two optimal transmission strategies in Fig. 5
with $M=2$. The upper part of Fig. 5 shows the save-transmit strategy,
i.e., the user is served by the grid-powered SCBS from time slot 1
to 4, and then by the EH-SCBS from time slot 5 to 10 (the critical
time slot index is 4). The lower part of Fig. 5 shows the greedy-transmit
solution, where the user is served by the EH-SCBS so long as it has
accumulated enough energy to support the transmit power. In both solutions,
the user is served by the grid-powered SCBS in four time slots, i.e.,
both consume the same amount of grid power.

These two solutions are typical transmission strategies for EH communication
systems. With a low complexity and a simple operation, surprisingly,
they are optimal for the considered problem. Though the optimality
may be lost in more general cases, they can still serve as heuristic
methods and provide low-complexity and sub-optimal solutions.

\subsection{Other Design Problems}

The above discussions of cell association shed light on the operation
issues in EH-SCNs, and provide some potential solutions. In general,
the design problems in EH-SCNs will be more challenging than in conventional
SCNs, and their unique characterizations, especially the impact of
energy profiles, should be taken into consideration. Such policies
as save-transmit and greedy-transmit can help develop efficient transmission
policies, which in certain cases can be shown to be optimal. There
are many other design problems to be addressed, including but not
limited to the following:
\begin{itemize}
\item \textbf{Sleep Control:} When taking the circuit power of an SCBS into
consideration, the energy efficiency of EH-SCNs can be effectively
improved by sleep control, i.e., to adaptively switch off some SCBSs.
\item \textbf{User Scheduling:} When an EH-SCBS is serving multiple users,
how to schedule these users is vital for the network performance.
As the available energy of each EH-SCBS accumulates over time, probably
the users with better channel conditions should be served earlier,
while the optimal policy requires further investigation.
\item \textbf{Channel Estimation:} Channel information is important for
wireless communications. However, due to the limited available energy
in EH-SCNs, the energy spent on channel estimation and data transmission
should be balanced. Moreover, it is also critical to decide when to
perform channel training based on the time-varying EH profile.
\end{itemize}

\section{Conclusions}


In this article, we conducted a comprehensive study of EH-SCNs, including
the feasibility analysis, network deployment investigation, and network
operation design. Among potential EH sources, we found that the combination
of solar and wind energy is a good candidate to power SCNs. To provide
network deployment guidelines for network deployment, three basic
tradeoffs between the network performance, EH-SCBS density, and grid
power consumption were investigated. For a given deployed EH-SCN,
 in order to optimize the network performance, special attention
was paid to the network operation designs in EH-SCNs. Throughout the
paper, distinctive challenges of EH-SCNs are highlighted, and novel
design methodologies are proposed. Open research problems are identified
which deserve unremitting efforts to promote faster, greener and more
flexible EH-SCNs.

{\scriptsize \bibliographystyle{IEEEtran}
\bibliography{mag_EH}
}{\scriptsize \par}

\section*{\newpage{}}

\section*{Biographies}

\begin{IEEEbiographynophoto}
{Yuyi Mao} [S'14] (ymaoac@ust.hk) received his B.Eng degree in Information and Communication Engineering from Zhejiang University, Hangzhou, China, in 2013. He is currently working towards the Ph.D. degree in the Department of Electronic and Computer Engineering at the Hong Kong University of Science and Technology, under the supervision of Prof. Khaled B. Letaief. His current research interests include energy harvesting cellular systems, cooperative systems, smart grid communications and stochastic optimization.
\end{IEEEbiographynophoto}
\begin{IEEEbiographynophoto}
{Yaming Luo} [S'11] (luoymhk@ust.hk) received his B.Eng. degree from the Department of Communication Engineering at Harbin Institute of Technology, Harbin, China, in 2010. He is currently working towards the Ph.D. degree in the Department of Electronic and Computer Engineering at the Hong Kong University of Science and Technology, under the supervision of Prof. Khaled B. Letaief. His current research interests include energy harvesting networks, relay systems, and green communications.
\end{IEEEbiographynophoto}
\begin{IEEEbiographynophoto}
{Jun Zhang} [M'10] (eejzhang@ust.hk) received the Ph.D. degree in Electrical and Computer Engineering from the University of Texas at Austin in 2009. He is currently a Research Assistant Professor in the Department of Electronic and Computer Engineering at the Hong Kong University of Science and Technology. Dr. Zhang co-authored the book \emph{Fundamentals of LTE} (Prentice-Hall, 2010). His research interests include wireless communications and networking, green communications, and signal processing.
\end{IEEEbiographynophoto}

\begin{IEEEbiographynophoto}
{Khaled B. Letaief} [S'85-M'86-SM'97-F'03] (eekhaled@ust.hk) received his Ph.D. from Purdue University. He is currently Chair Professor and Dean of Engineering at HKUST. He is an internationally recognized leader in wireless communications with over 500 papers and 15 patents. He is founding Editor-in-Chief of \emph{IEEE Transactions on Wireless Communications} and recipient of many honors including 2009 \emph{IEEE Marconi Prize Award in Wireless Communications} and 12 IEEE Best Paper Awards. He is an \emph{IEEE Fellow} and \emph{ISI Highly Cited Researcher}.
\end{IEEEbiographynophoto}

\section*{\newpage{}}

\begin{center}
\begin{figure}
\begin{centering}
\subfloat[Short time horizon.]{\begin{centering}
\includegraphics[width=8cm]{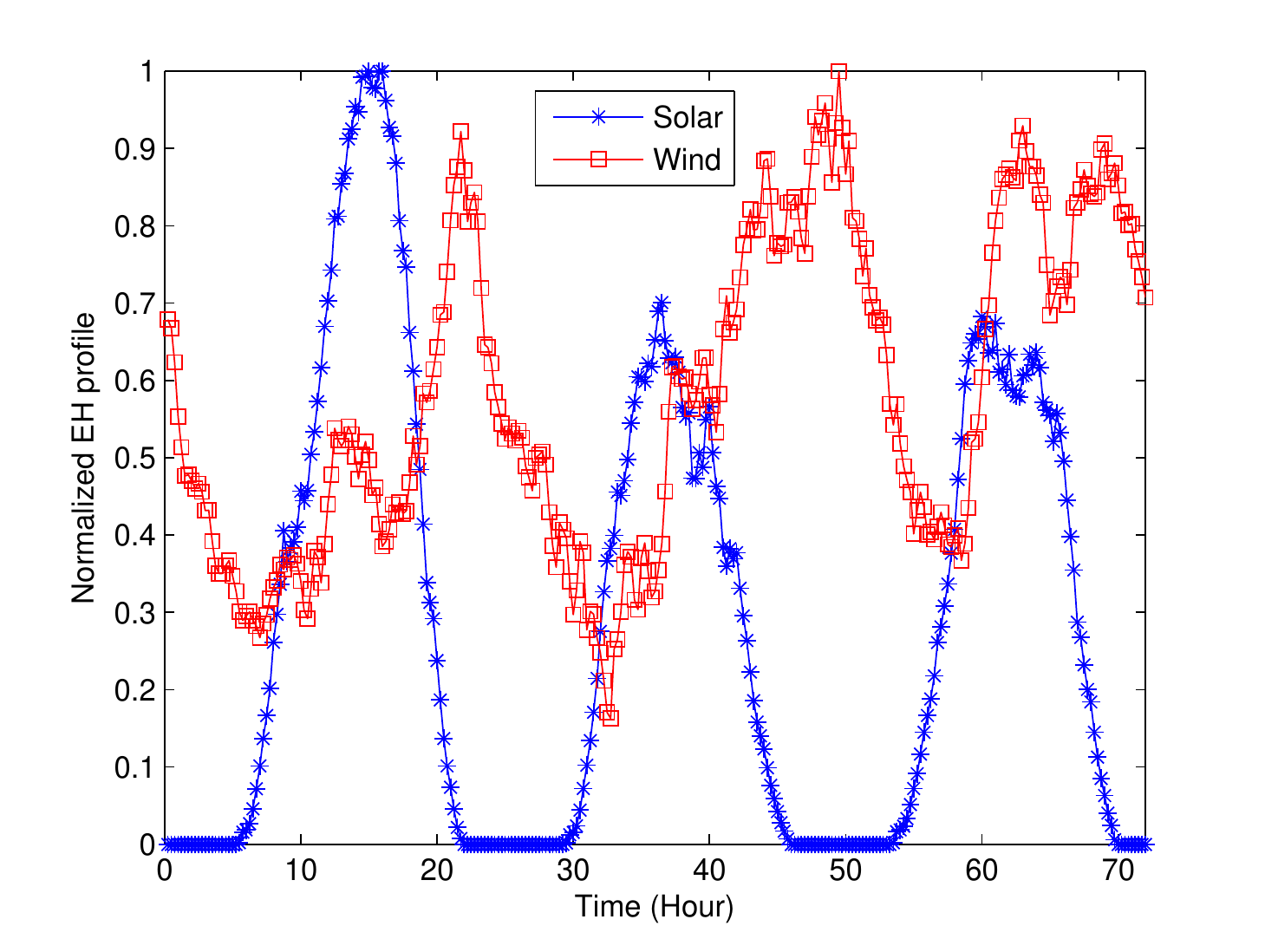}
\par\end{centering}

} \subfloat[Long time horizon.]{\begin{centering}
\includegraphics[width=8cm]{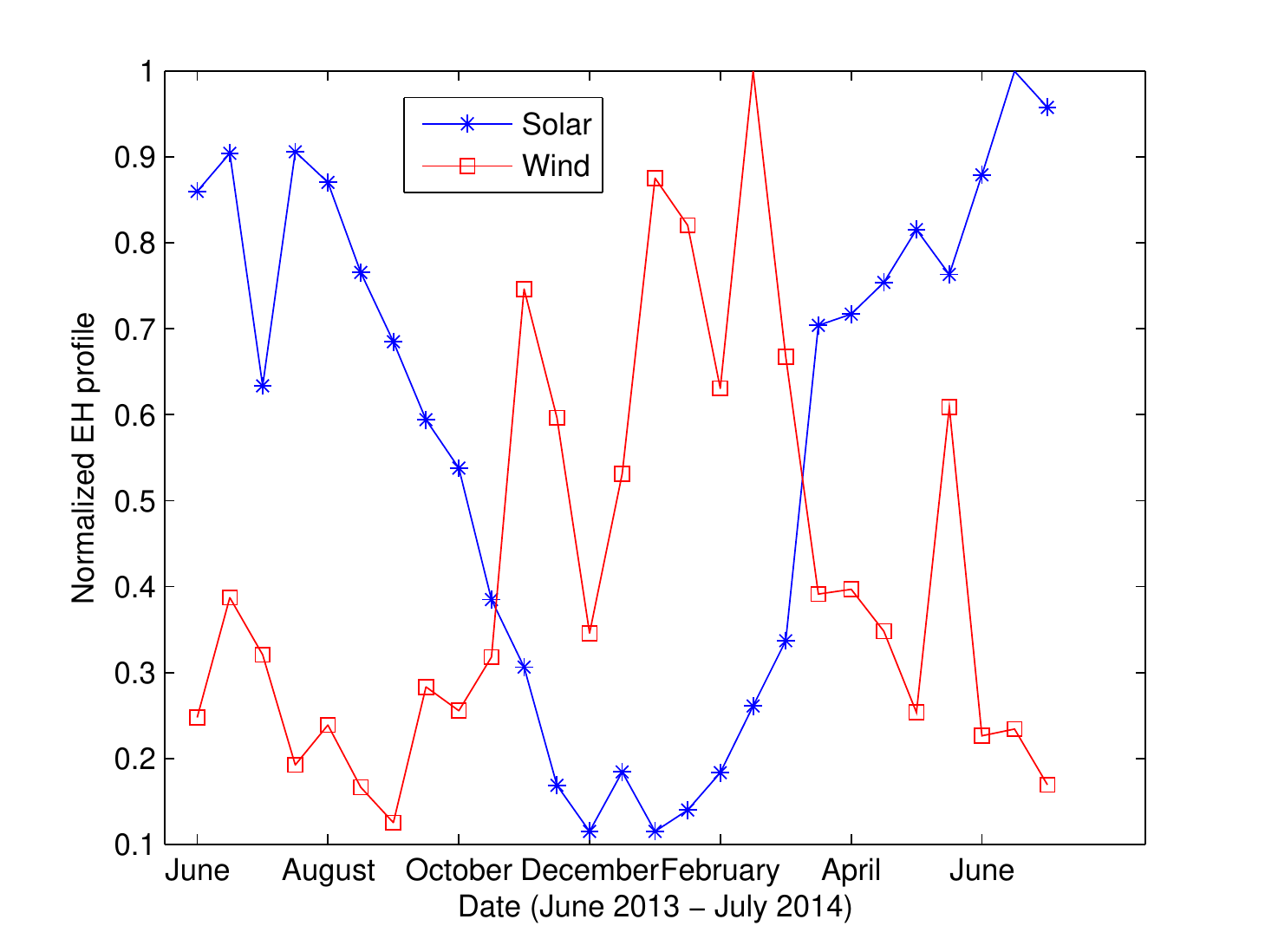}
\par\end{centering}

}
\par\end{centering}

\caption{Normalized solar and wind energy profiles.}
\end{figure}

\par\end{center}

\section*{\newpage{}}

\begin{center}
\begin{figure}[h]
\begin{centering}
\label{System Model Fig}
    \includegraphics[width=14cm]{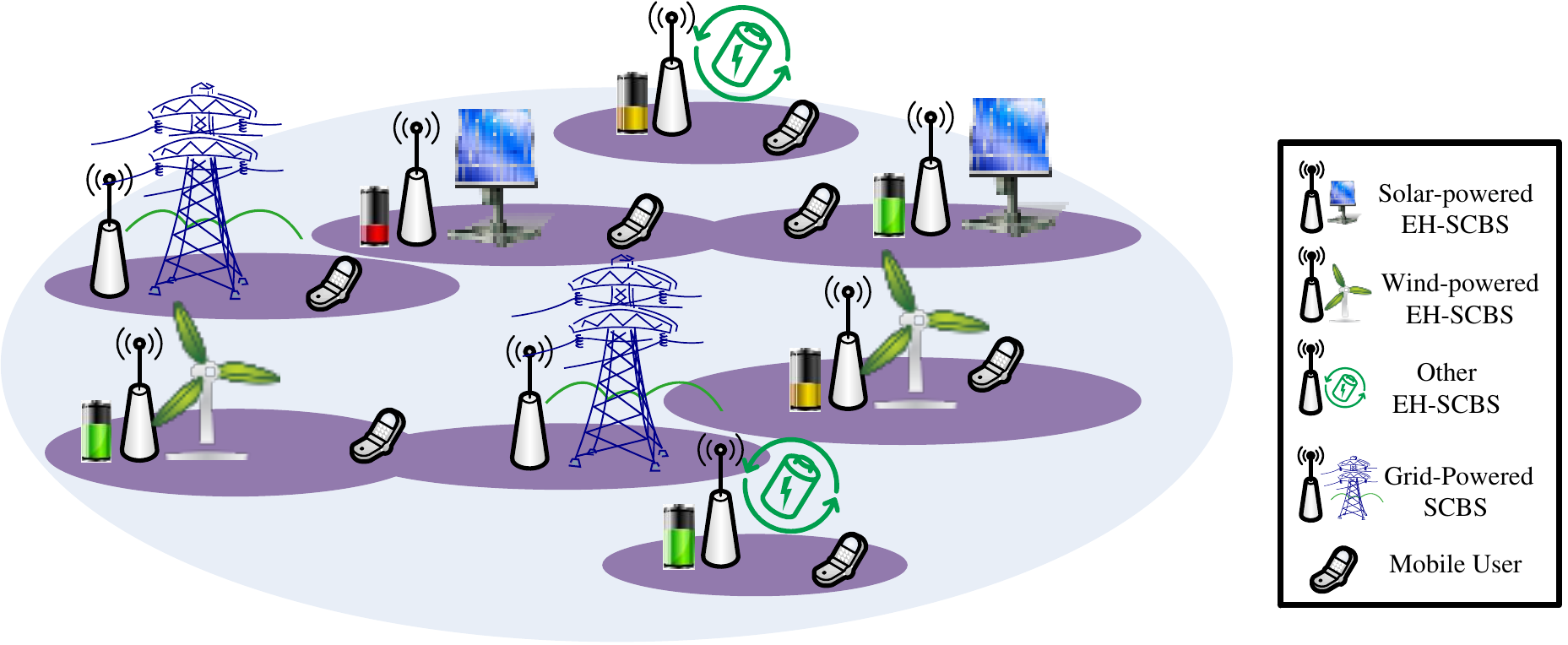}
\par\end{centering}

\caption{A sample EH-SCN.}
\end{figure}

\par\end{center}

\section*{\newpage{}}

\begin{center}
\begin{figure}
\begin{centering}
\subfloat[Outage probability vs. SCBS density.]{\begin{centering}
\includegraphics[width=8cm]{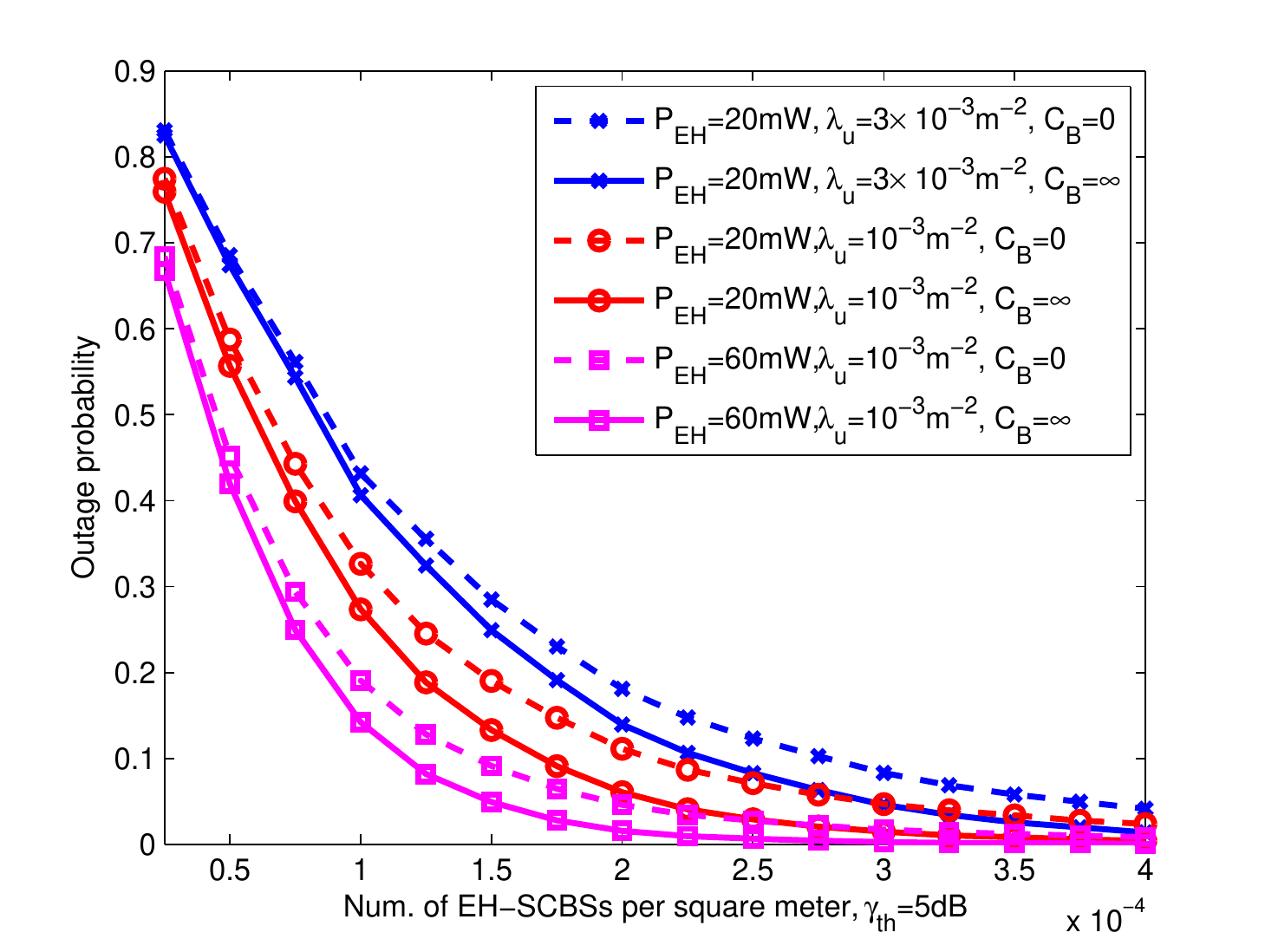}
\par\end{centering}

} \subfloat[Grid power consumption vs. SCBS density.]{\begin{centering}
\includegraphics[width=8cm]{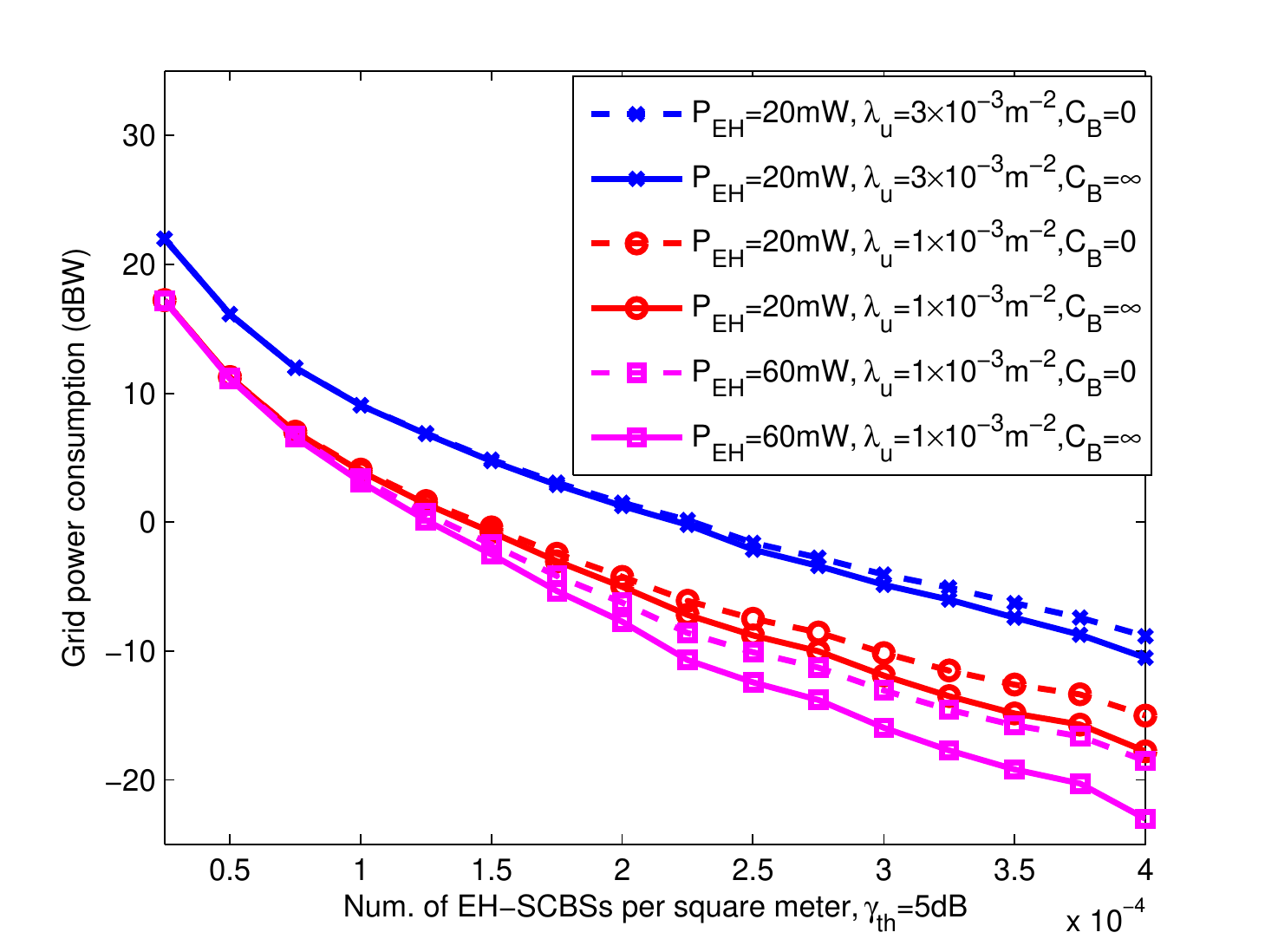}
\par\end{centering}

}
\par\end{centering}

\begin{centering}
\subfloat[Outage probability vs. grid power consumption.]{\begin{centering}
\includegraphics[width=8cm]{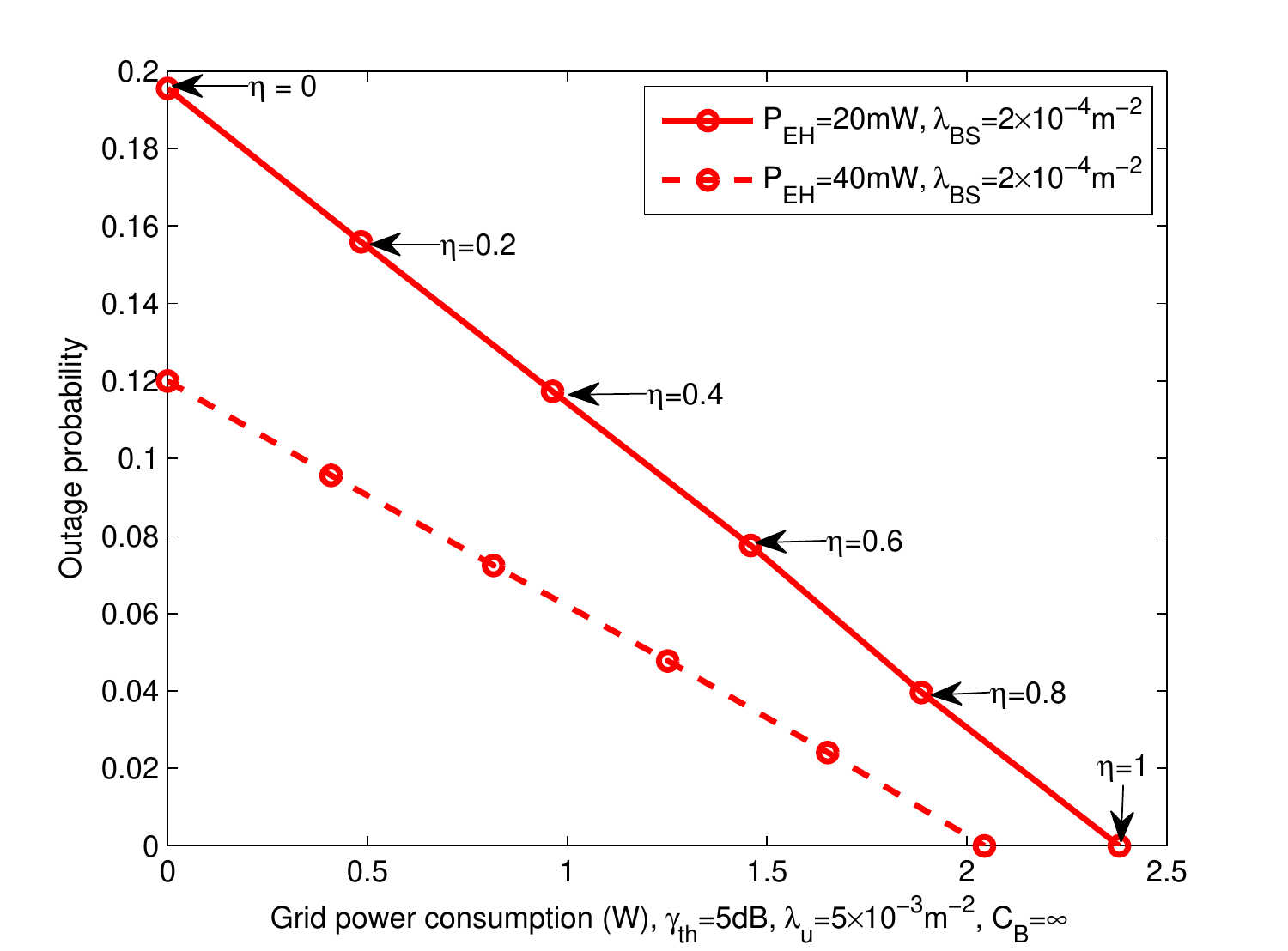}
\par\end{centering}

}
\par\end{centering}

\caption{Basic tradeoffs in the EH-SCNs.}
\end{figure}

\par\end{center}

\section*{\newpage{}}

\begin{center}
\begin{figure}[h]
\begin{centering}
\label{System Model Fig-1} %\includegraphics[width=8cm,height=4cm]{sys_archiet.eps}
    \includegraphics[width=14cm]{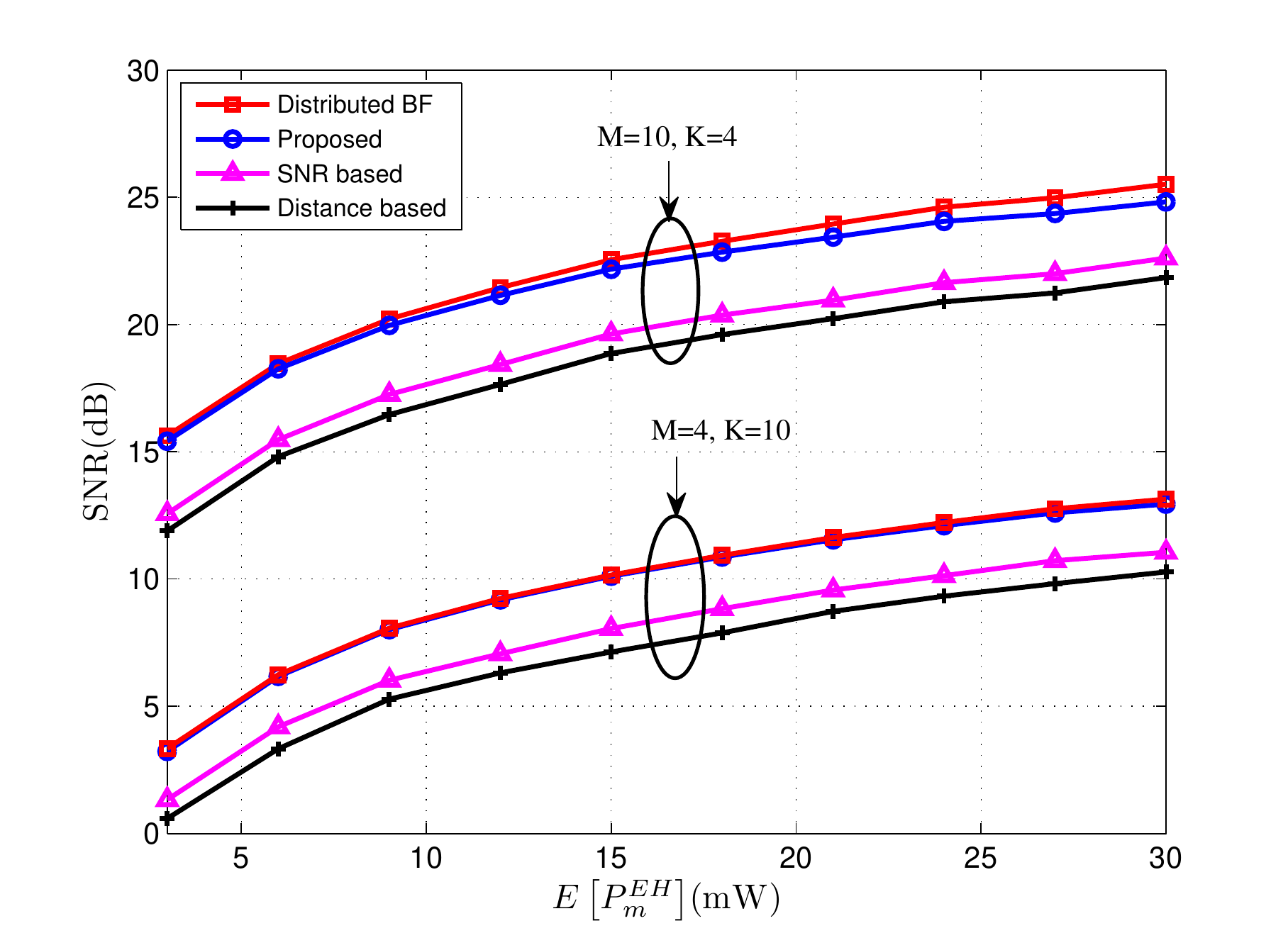}
\par\end{centering}

\caption{Comparisons of different power assignment and cell association policies.}
\end{figure}

\par\end{center}

\section*{\newpage{}}

\begin{center}
\begin{figure}[h]
\begin{centering}
\label{System Model Fig-1-1} %\includegraphics[width=8cm,height=4cm]{sys_archiet.eps}
    \includegraphics[width=14cm]{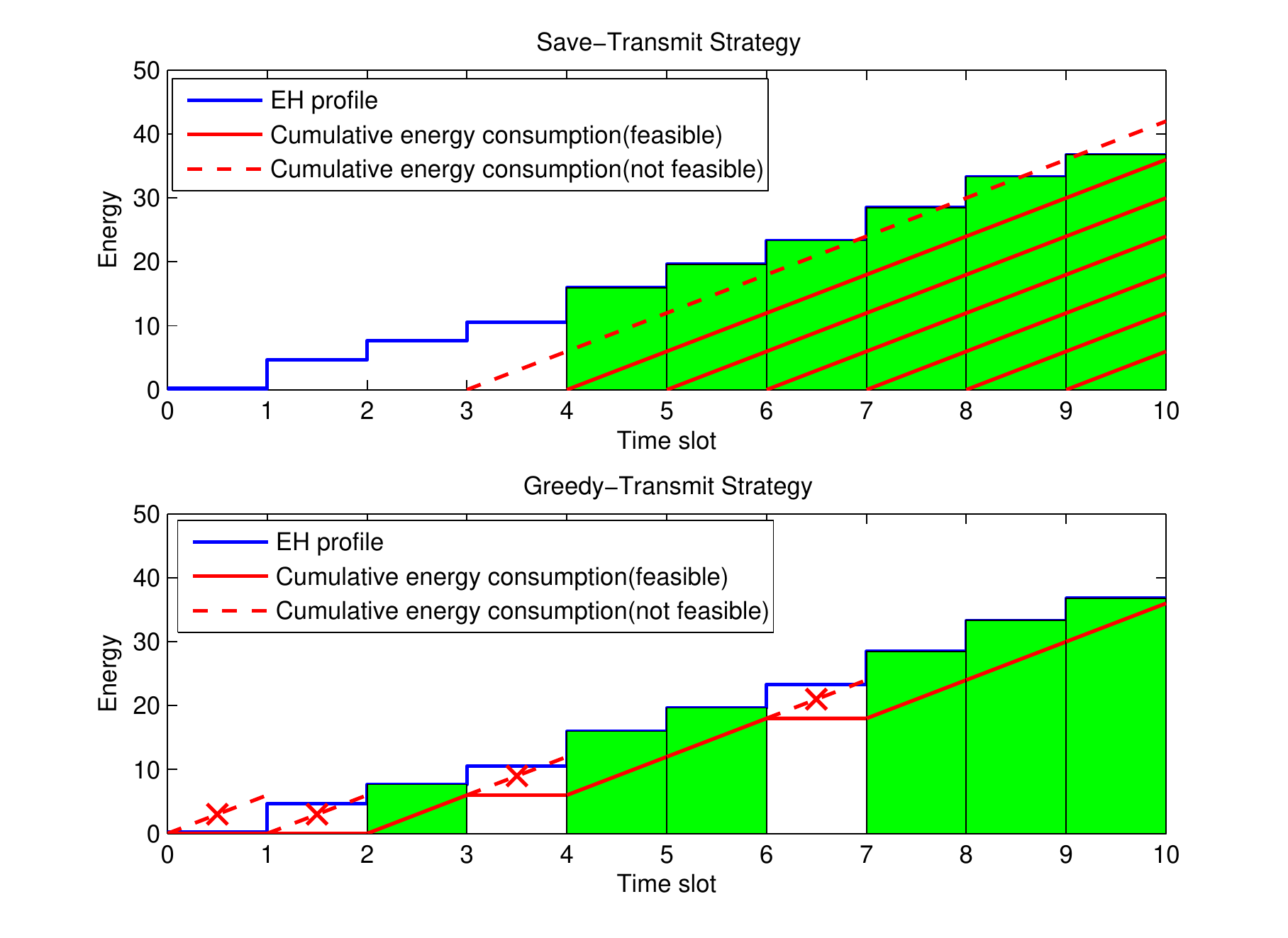}
\par\end{centering}

\caption{Illustration of the Save-Transmit and Greedy-Transmit strategies (the
user is served by the EH-SCBS in the shaded time slots, and the slope
of the curve represents the transmit power).}
\end{figure}

\par\end{center}
\end{document}